\documentclass[fleqn,twoside]{article}
\usepackage{espcrc2}
\usepackage{amssymb}

\title{Benchmarking computer platforms for lattice QCD applications}

\author{%
M.~Hasenbusch\address[NIC]{NIC/DESY-Zeuthen, 15738 Zeuthen, Germany},
K.~Jansen\addressmark[NIC],
D.~Pleiter\addressmark[NIC],
H.~St\"uben\address{Konrad-Zuse-Zentrum f\"ur Informationstechnik Berlin, 
14195 Berlin, Germany},
P.~Wegner\address{DESY-Zeuthen, 15738 Zeuthen, Germany},
T.~Wettig\address{Department of Physics, Yale University, 
New Haven, CT 06520--8120, USA},
H.~Wittig\address{DESY, 22603 Hamburg, Germany}
}

\newcommand{\csw}{c_{\rm{sw}}}
\newcommand{\daxpy}{\textsl{\textsf{daxpy}}}
\newcommand{\zaxpy}{\textsl{\textsf{zaxpy}}}

\newlength{\dWidth}    
\newcommand{\D}{\makebox[\dWidth]{}}

\begin{document}

\begin{abstract}
We define a benchmark suite for lattice QCD and report on benchmark
results from several computer platforms.  The platforms considered are
apeNEXT, CRAY T3E, Hitachi SR8000, IBM p690, PC-Clusters, and QCDOC.
\vspace*{-1mm}
\end{abstract}

\maketitle

\section{INTRODUCTION}

Simulations of lattice QCD require powerful computers.
We have benchmarked computers that are under consideration by the German
Lattice Forum (LATFOR) \cite{latfor} to realize the future
physics program.
The machines we have tested fall into three
categories: (1) machines that are custom-designed for lattice QCD
(apeNEXT and QCDOC), (2) PC-clusters, and (3) commercial
supercomputers (CRAY, Hitachi, IBM).

\vspace*{-1mm}
\section{COMPUTER PLATFORMS}

\subsection{apeNEXT}

The apeNEXT project~\cite{ape} was initiated with the goal to build
custom-designed computers with a peak performance of more than 5
TFlops and a 
sustained efficiency of about 50\% for key lattice gauge theory kernels.
apeNEXT machines should be suitable for both large-scale simulations
with dynamical fermions and quenched calculations on very large
lattices.  The apeNEXT processor is a 64-bit architecture with an
arithmetic unit that can at every clock cycle perform the APE normal
operation $a \times b + c$, where $a$, $b$, and $c$ are IEEE 128-bit
complex numbers.  The apeNEXT processors
have a very large register file of 256 (64+64)-bit registers.
On-chip network devices connect the nodes by a three-dimensional
network.

\subsection{QCDOC}

QCDOC (``QCD on a Chip'')~\cite{qcdoc} is a massively parallel
computer optimized
for lattice QCD, developed by a collaboration of Columbia University,
UKQCD, the RIKEN-BNL Research Center, and IBM.  Individual nodes are
based on an application-specific integrated circuit (ASIC) which
combines IBM's system-on-a-chip technology (including a PowerPC 440
CPU core, a 64-bit FPU, and 4 MB on-chip memory) with custom-designed
communications hardware.  The nodes communicate via nearest-neighbor
connections in six dimensions.  The low network latency and built-in
hardware assistance for global sums enable \mbox{QCDOC} to concentrate
computing power in the TFlops range on a single QCD problem.

\subsection{PC-cluster}

In recent years, commodity-off-the-shelf (COTS) Linux cluster computers
have become cost-efficient, general-purpose, high-performance computing
devices.  QCD simulations on cluster computers can be boosted
considerably by exploiting the SIMD and data prefetch functionality of
Intel Pentium processors via SSE/SSE2 instructions  
by means of
assembler coding.  
The benchmarked PC-cluster has 1.7 GHz Xeon Pentium 4 CPUs with 1
GB of Rambus memory. The nodes communicate via a Myrinet2000
interconnect.

\subsection{CRAY T3E-900}

The CRAY T3E is a classic massively parallel computer.  It has single
CPU nodes and a three-dimensional torus network.  The T3E
architecture is rather well balanced.  Therefore, the overall performance
of parallel applications scales to much higher numbers of CPUs than on
machines that were built later.  The peak performance of a T3E-900 is 900
MFlops per CPU, the network latency is 1~$\mu$s, and the bidirectional
network bandwidth is 350 MByte/s.

\subsection{Hitachi SR8000-F1}

The Hitachi SR8000 is a parallel computer with shared memory nodes.
Each node has 8 CPUs.  The key features of the CPUs are the high memory
bandwidth and the availability of 160 floating point registers.  These
features are accompanied by pseudo-vectorization, an intelligent
pre-fetch mechanism that allows to overlap computation and fetching
data from memory.  Pseudo-vectorization is done by the compiler.  The
peak performance of an SR8000 CPU is 1500 MFlops, the network latency
is 19~$\mu$s, and the bidirectional bandwidth between nodes is 950
MByte/s.

\subsection{IBM p690-Turbo}

The IBM p690 is a cluster of shared memory nodes.  Its CPUs (and
nodes) have the highest peak performance of the machines considered
but only a relatively slow network.  In order to increase the
bandwidth of the interconnect people divide the 32-CPU nodes into
8-CPU nodes.  This increases the bandwidth per CPU by a factor of 4.
The performance depends to a large extent on the configuration
of the machine. For benchmarking this architecture it has also to be
taken into account that the performance drops by a factor of 3--5
when using all CPUs instead of only one.  The peak performance
of a Power4 CPU is 5400 MFlops, the network latency (of the so-called
\emph{colony} network) is 20~$\mu$s, and the bidirectional bandwidth
between nodes is 450 MByte/s.

\begin{table*}[t]
\caption{Benchmark results in MFlops per CPU. All numbers refer to
64bit floating point arithmetic.
\textit{Italic} numbers indicate that communications overhead has been
included. 
Further details are given in the text.}
\begin{tabular*}{\textwidth}{l%
@{\extracolsep{\fill}}c%
@{\extracolsep{\fill}}c%
@{\extracolsep{\fill}}c%
@{\extracolsep{\fill}}c%
@{\extracolsep{\fill}}c%
@{\extracolsep{\fill}}c}
\hline
            & apeNEXT & QCDOC & PC-Cluster & CRAY & Hitachi & IBM \\
 Peak [MFlops]     & 1600 & 1000 & 3400 & 900  & 1500 & 5200 \\
\hline\\[-3mm]
 $H_{\rm eo}\,\phi$&\D\textit{894} &\D\textit{535} &\D930 & 
                      \textit{101} &\D\textit{632} & \textit{299} 
\\[\smallskipamount]
 $(\psi,\phi)$     &\D656 &\D450 &\D530 & 148  &\D680 & 303 \\
 $||\psi||^2$      &\D592 &\D384 &\D510 &\D98  &\D789 & 187 \\
 \zaxpy            &\D464 &\D450 &\D358 & 114  &\D479 & 234 \\
 \daxpy            &\D116 &\D190 &\D183 &\D57  &\D241 & 115 
\\[\smallskipamount]
 $U$*$\phi$        & 1264 &\D780 &\D307 & 104  &\D811 & 261 \\
 $U$*$V$           & 1040 &\D800 &\D763 & 118  & 1182 & 413 
\\[\smallskipamount]
 $T\phi$           & 1136 &\D790 &\D800 & 111  & 1137 & 608 \\
\hline
\end{tabular*}
\vspace*{-4mm}
\end{table*}

\section{BENCHMARK SUITE}

In this contribution we concentrate on one particular application:
large-scale simulations of dynamical Wilson fermions with $O(a)$-improvement
on lattices of size $V=32^3\times 64$.
We assume that these simulations are performed
using the Hybrid Monte Carlo algorithm \cite{HMC} or the Polynomial
Hybrid Monte Carlo algorithm \cite{PHMC}, as  was done in
simulations with dynamical fermions in recent years
\cite{lat}.

The most time-consuming operation is the fermion matrix
multiplication. We denote the fermion matrix by $M[U] = T[U] - H[U]$,
where $H$ is the Wilson hopping term
\begin{eqnarray}
\lefteqn{H[U]_{xy} = \kappa \sum_{\mu} \big\{
      (1-\gamma_{\mu})\,U_{\mu}(x)\,\delta_{x+\hat\mu,y}} \nonumber \\
  & \hspace*{20mm} + (1+\gamma_{\mu})\,U_{\mu}^{\dag}(x-\hat\mu)\,
      \delta_{x-\hat \mu,y} \big\}
\end{eqnarray}
and $T$ is the clover term $T[U] =
1-{\textstyle\frac{i}{2}}\kappa\csw\,F_{\mu\nu}\sigma_{\mu\nu}$.
Here we only consider the even-odd
preconditioned version $\psi = H_{\rm eo}\, \phi$.

Basic operations of linear algebra are needed in iterative solvers.
We have considered the scalar product
\begin{equation}
(\psi,\phi) = 
       \sum_{x=1}^V\sum_{i=1}^3\sum_{\alpha=1}^4
       \psi^*_{i,\alpha}(x)\phi_{i,\alpha}(x)\,,
\end{equation}
the vector norm
\begin{equation}
||\psi||^2  =
       \sum_{x=1}^V\sum_{i=1}^3\sum_{\alpha=1}^4
       |\psi_{i,\alpha}(x)|^2\,,
\end{equation}
the \zaxpy\ operation
\begin{equation}
\psi_{i,\alpha}(x)
           \leftarrow \psi_{i,\alpha}(x)+c\,\phi_{i,\alpha}(x),\quad
           c\in\mathbb{C}\,,
\end{equation}
and the \daxpy\ operation
\begin{equation}
\psi_{i,\alpha}(x)
           \leftarrow \psi_{i,\alpha}(x)+r\,\phi_{i,\alpha}(x),\quad
           r\in\mathbb{R}\,.
\end{equation}

Two basic operations involving link variables are part of our
benchmark, the multiplication of an $SU(3)$ matrix by a vector
\begin{equation}
\psi = U*\phi; \quad
  \psi_i=\sum_{j=1}^3 U_{ij}\phi_j
\end{equation}
and the multiplication of two $SU(3)$ matrices
\begin{equation}
W = U*V;    \quad
   W_{ij}=\sum_{k=1}^3 U_{ik}V_{kj}\,.
\end{equation}

Finally, the benchmark contains the basic operations involving the
clover term, $\psi = T \phi$ and $\psi = T^{-1} \phi$. These were
implemented with $6 \times 6$ block matrices,
\begin{equation}
\psi =
\frac{1}{2}
\left(
\begin{array}{rr} 1 & -1 \\ 1 & 1 \\ \end{array}
\right)
\left(
\begin{array}{cc} \Box & 0 \\ 0 & \Box \\ \end{array}
\right)
\left(
\begin{array}{rr} 1 & 1 \\ -1 & 1 \\ \end{array}
\right)
\phi\:.
\end{equation}

\section{BENCHMARK RESULTS}

Our benchmark results are listed in Table~1.
The values for apeNEXT and QCDOC were
obtained from cycle-accurate simulations of the forthcoming hardware.
All the other performance numbers were measured on existing machines.

On apeNEXT and QCDOC the hopping term was benchmarked by distributing
the problem over the maximum number of nodes for the given problem
size.  For the PC-cluster, where a C code with SSE/SSE2 instructions
based on the benchmark program of M.~L\"{u}scher~\cite{luescher} has
been used, only single-node numbers (for $V=16^4$) are quoted because
there is still some debate over which network to use (e.g., Myrinet,
Infiniband, Gbit-Ethernet).
The commercial machines have been benchmarked using 256, 64 and 64
CPUs on CRAY, Hitachi and IBM, respectively.
We used the Fortran90 production code of the QCDSF collaboration
that is parallelized with MPI and OpenMP.
For the linear algebra routines we used Fortran loops on the
Hitachi and the vendors' high-performance libraries
on CRAY and IBM.

In case of the scalar product and the vector norm we quote
only the single-processor performance, since the performance
including the global sum depends on the number of nodes.
We estimated the overhead for computing the global sum on some
platforms, since it
will affect scalability of the considered application when going to
a very large number of nodes:

\smallskip
\noindent\hspace*{-6pt}
\begin{tabular}{ll}
apeNEXT    & 5.2 $\mu\mbox{s}$ on 4 $\times$ 8 $\times$ 8 = 256 CPUs \\
QCDOC      & 10 (15) $\mu\mbox{s}$ on 4.096 (32.768) CPUs \\
PC-cluster & 138 (166) $\mu\mbox{s}$ on 256 (1.024) CPUs
\end{tabular}
\vspace*{-0.5mm}

\vspace*{-2mm}
\section{CONCLUSIONS}

We presented a selection of benchmarks relevant for doing large-scale
simulations of QCD with dynamical fermions and provided 
initial benchmark results for a range of platforms. A more detailed
comparison of these platforms in terms of
price/performance ratio, hardware reliability, software support, etc.
is beyond the scope of this contribution.
These questions will be addressed in a future publication.

\end{document}